\documentstyle[11pt]{article}
\begin{document}

\large

 {\huge{\bf
 \centerline {The  Structure of the Extreme}
 \centerline {Schwarzschild--de Sitter  Space-time}}}
 \vspace{15mm}
 \centerline {\bf J.  Podolsk\' y}
 \vspace{15mm}
 {\it
 \centerline {Department of Theoretical Physics,}
 \centerline {Charles University, Faculty of Mathematics and Physics,}
 \centerline {V Hole\v sovi\v ck\'ach 2, 180 00 Prague 8, Czech Republic}}
 \vspace{8mm}
 {\footnotesize
 \centerline {Electronic address:
              podolsky@mbox.troja.mff.cuni.cz}}

 \vspace{10mm}
\begin{abstract}
The extreme Schwarzschild--de Sitter space-time is a spherically
symmetric solution of Einstein's equations with a cosmological
constant $\Lambda$ and mass parameter $m>0$ which is characterized
by the condition that $9\Lambda m^2=1$.
The global structure of this space-time is here analyzed in
detail. Conformal and embedding diagrams are constructed, and
synchronous coordinates which are suitable for a discussion of
the cosmic no-hair conjecture are presented. The permitted
geodesic motions are also analyzed. By a careful investigation of
the geodesics and the equations of geodesic
deviation, it is shown that specific families
of observers escape from falling into the singularity and
approach nonsingular asymptotic regions which are represented by special
``points'' in the complete conformal diagram. The redshift of
signals emitted by particles which fall into the singularity, as
detected by those observers which escape, is also calculated.

\vspace{10mm}
\vspace{2mm}

\noindent
PACS NUMBERS: 04.20.Jb, 04.70.Bw

\end{abstract}

\newpage
\section {INTRODUCTION}

Black holes in (anti -) de Sitter space-time have attracted
increased attention in the last two decades. Various aspects of
the well-known Schwarzschild--de Sitter solution \cite{Kottler},
\cite{KSMH}
(a spherically symmetric vacuum solution with a cosmological constant
$\Lambda>0$) containing both the black-hole and cosmological
horizons have been studied. In particular its global
structure \cite{GH}-\cite{CuLa}, geodesic motion \cite{How}-\cite{Jak},
quantum effects \cite{BoHa}-\cite{ON},
or behavior of fields in this background \cite{KP}-\cite{BCKL}
have been clarified.
Other members of a general Kerr-Newman-de~Sitter family of solutions
\cite{Car} have also been investigated including the extreme cases
\cite{GH}-\cite{Ge},\cite{NMNO}-\cite{Stu},
\cite{BCKL}-\cite{WSY}. Many papers have concentrated
on the interesting problem of the stability of the Cauchy horizon.
There has been a growing body of evidence that (in some cases)
the Cauchy horizon is classically stable to linear and even
non-linear time dependent perturbations. Thus, black hole-de
Sitter space-times can serve as counter-examples of the strong
censorship hypothesis (see e.g. \cite{Cham} for a detailed
discussion and number of references).

Recently, exact multi-black-hole  solutions in asymptotically de Sitter
universe have been discovered and discussed \cite{KaTr}.
These space-times describe systems of an arbitrary number of
``extremally charged'' black holes. Their specific properties have been
investigated in subsequent papers \cite{BHKT}-\cite{INSH}.
In particular, collisions of extreme black holes and the cosmic
censorship hypothesis have been studied within this framework.

There are also exact models of black-hole formation in the
presence of a cosmological constant $\Lambda$. In  \cite{BiPo1}
it was shown that the Robinson-Trautman type $II$ spacetimes with
$\Lambda$  converge asymptotically  to a corresponding spherically symmetric
Schwarzschild--(anti-) de~Sitter solution for large retarded times
(i.e., near the future event horizon) with the initial
``perturbation'' being radiated away in gravitational waves.
In general, the extension of these space-times across the event horizon
can only be made with a finite degree of smoothness. However, in
the extreme case the horizon is smooth but nonanalytic \cite{BiPo2}.

Some of the results in the above mentioned papers suggest that
some properties of extreme black holes are often qualitatively different from
those of generic ones. This is the primary motivation of our work
in which we concentrate on the solution describing
{\it extreme} Schwarzschild--de Sitter black holes;
we investigate and systematically summarize its properties.

In the  next section we introduce the extreme Schwarzschild--de Sitter
space-time. In Section 3, we present Kruskal-type null
coordinates and we rigorously construct the conformal diagram using
a proper conformal factor. Sections 4 and 5 are devoted to
construction of the embedding diagram and synchronous coordinates.
In Section 6, a complete investigation of all geodesics (null, timelike,
and spacelike) in the extreme Schwarzschild--de Sitter space-time
is presented. In Section 7, relative motions described by the
equation of geodesic deviation are studied. Section 8 is devoted
to an analysis of interesting ``asymptotic points'' in the
space-time, and in Section 9 the redshift viewed by observers approaching
these points is calculated.
The results are summarized in the concluding Section 10.

\section {THE EXTREME SCHWARZSCHILD--DE SITTER SPACE-TIME}

The metric  of a generic Schwarzschild--de Sitter space-time in
standard coordinates is
\begin{equation}
ds^2 = -\Phi\, dt^2+\Phi^{-1}\, dr^2+
      r^2(d\theta^2+\sin^2\theta d\varphi^2) \ , \label{E2.1}
\end{equation}
where $\Phi (r) = 1-2m/r-(\Lambda/3)\,r^2$ with
$\Lambda>0$ and $m>0$. For $0<9\Lambda m^2<1$ there exist
two positive roots  $r_+$ and $r_{++}$ of $\Phi(r)$
such that $0<2m<r_+<3m<r_{++}$. The root
$r_+=(2/\sqrt\Lambda)\cos(\alpha/3+4\pi/3)$,
with $\cos\alpha=-3m\sqrt\Lambda$, describes the black-hole event
horizon, and the root $r_{++}=(2/\sqrt\Lambda)\cos(\alpha/3)$
localizes the cosmological event horizon. Such space-times have
been discussed in detail for example in \cite{GH}-\cite{Jak} and
elsewhere.

As $\Lambda$ approaches its extremal value,
$\Lambda\rightarrow1/9m^2$, the position of the black-hole horizon $r_+$
monotonically increases and the cosmological horizon $r_{++}$
decreases to the common value $3m$. In this paper we analyze
this extreme case  of the Schwarzschild--de Sitter space-time
which is characterized by the condition $9\Lambda m^2=1$ (related
to the Nariai solution \cite{Nar}, see e.g. \cite{BoHa}). In this
case there exists only one degenerate ``double'' Killing horizon
at $r=3m$; this can be seen from the corresponding form of $\Phi(r)$,
\begin{equation}
\Phi (r) = -\frac{1}{27m^2\,r} (r-3m)^2(r+6m) \ . \label{E2.2}
\end{equation}
The surface gravity of the horizon is $\kappa=0$. Also,
$\Phi \le 0$ everywhere, so that $r$ is a time coordinate,
$t$ is a spatial coordinate, and there is no static region
in the extreme Schwarzschild--de Sitter space-time.

\section {GLOBAL STRUCTURE}

The global structure of the extreme Schwarzschild--de Sitter
space-time has already been described, for example in
\cite{Lake}, \cite{Ge} and elsewhere. However, as far as we know,
the corresponding conformal diagram has not yet been constructed rigorously.
For example, in \cite{Lake} double-null coordinates
were found, but the ``compactifying'' conformal transformation was
not given, so that the diagram was only
schematic. In \cite{Ge} the conformal
transformation was given but it cannot be applied in both
regions above and below the horizon simultaneously.
Here we overcome these obstacles and present an
exactly constructed conformal digram together with an explicit form of
the conformal factor.

Introducing  the Kruskal-type null coordinates $\hat u$, $\hat v$ by
\begin{eqnarray}
&&u=\delta\cot\hat u\ ,   \nonumber\\
&&v=\delta\tan\hat v\ , \label{E3.1}
\end{eqnarray}
where $\delta=-m(3-2\ln2)<0$, $u=t-r^*$, $v=t+r^*$, and
the ``tortoise'' coordinate
\begin{equation}
r^*=\int\frac{dr}{\Phi}=
\frac{9m^2}{r-3m}+2m\ln\left|\frac{r+6m}{r-3m}\right|\ ,
                                                  \label{E3.2}
\end{equation}
(an additive constant was chosen such that $r^*\rightarrow0$
at $r\rightarrow\infty$), the metric of the extreme
Schwarzschild--de Sitter space-time can be written in the form
\begin{equation}
ds^2=-\frac{\delta^2}{27m^2 r}
  \frac{(r+6m)(r-3m)^2}{\sin^2\hat u\,\cos^2\hat v}d\hat u d\hat v
  +r^2 (d\theta^2+\sin^2\theta d\varphi^2) \ .  \label{E3.3}
\end{equation}
The conformal diagram which can easily be obtained from Eqs.
(\ref{E3.1}), (\ref{E3.2})
is drawn in Fig.~\ref{Fig.1}  --- the space-time
may represent either extreme  black holes (Fig.~\ref{Fig.1} (a))
or white holes (Fig.~\ref{Fig.1} (b))
in the de Sitter universes. These two possibilities are connected
by a simple reflection $\hat u\rightarrow-\hat u$,
$\hat v\rightarrow-\hat v$; the parametrization (\ref{E3.1})
gives the white-hole case. The metric  (\ref{E3.3})
is regular even on the horizon given by $r=3m$ since
\begin{equation}
\lim_{r\rightarrow 3m}\left|\frac{r-3m}{\sin\hat u}\right|=
\lim_{r\rightarrow 3m}\left|\frac{r-3m}{\cos\hat v}\right|=
\frac{18m^2}{-\delta}                  \label{E3.4}
\end{equation}
for all finite fixed $u$ and $v$, respectively. The causal structure is
evident. Any timelike geodesic observer falling from the region
$r>3m$ (or the infinity ${\cal J}^- $ given by
$r=\infty$) in the black-hole space-time will either cross
the horizon $r=3m$ and reach the singularity at $r=0$, or escape to
one of the ``asymptotic points'' ${\cal P}$ given by $u=-\infty$, $v=+\infty$.
In the white-hole case the observers ``emanate'' from the singularity
at $r=0$ (or from the points ${\cal P}$) and, after crossing the horizon,
they reach the future infinity ${\cal J}^+$  ($r=\infty$) or the asymptotic
points ${\cal Q}$ (given by $u=+\infty$, $v=-\infty$).

Now we shall demonstrate that the infinity ${\cal J}^+$ given by
$r=\infty$, i.e. $\hat u+\hat v=\frac{\pi}{2}$ in the white-hole
space-time (the proof for ${\cal J}^-$ in the black-hole case
is analogous), is smooth and ``de Sitter-like''. First,
we introduce null coordinates $U$, $V$ by
\begin{eqnarray}
&&\cot\hat u=\frac{\sqrt{27}\,m}{-\delta}
  \ln\left|\cot\frac{U}{2}\right|\ ,   \nonumber\\
&&\tan\hat v=\frac{\sqrt{27}\,m}{-\delta}
  \ln\left|\cot\left(\frac{\pi}{4}-\frac{V}{2}\right)\right|\ , \label{E3.4b}
\end{eqnarray}
in which the metric takes the form
\begin{equation}
ds^2=-\left(1+\frac{6m}{r}\right)\frac{(r-3m)^2}{\sin U\,\cos V}dU dV
  +r^2 (d\theta^2+\sin^2\theta d\varphi^2) \ ,      \label{E3.4c}
\end{equation}
where $r=r(r^*)$ according to (\ref{E3.2}) and
$r^*=\frac{\sqrt{27}}{2}m(\ln|\cot\frac{U}{2}|-
\ln|\cot(\frac{\pi}{4}-\frac{V}{2})|)$.
The relation $(\hat u, \hat v)\leftrightarrow(U,V)$ given by (\ref{E3.4b})
is a one-to-one correspondence. In particular, $\hat u\in(0,\pi)$
is uniformly mapped to $U\in(0,\pi)$ and similarly
$\hat v\in(-\frac{\pi}{2},\frac{\pi}{2})$ is mapped to
$V\in(-\frac{\pi}{2},\frac{\pi}{2})$.
(Moreover, by introducing $\tilde v=\frac{\pi}{2}-\hat v$ and
$\tilde V=\frac{\pi}{2}-V$ we observe that
$\tilde v(\tilde V)$ is the same function as $\hat u(U)$.)
In these coordinates the scri ${\cal J}^+$ at
$r=\infty$ is given by $U+V=\frac{\pi}{2}$.

Now, choosing a conformal factor
\begin{equation}
\Omega^2=\frac{r}{r+6m}\frac{\sin U\cos V}{(r-3m)^2}
   \ ,                                          \label{E3.5}
\end{equation}
we can write the conformal (unphysical) metric as
\begin{equation}
d\hat s^2=\Omega^2 ds^2= - d U d V
  +\frac{r}{r+6m}\left(\frac{r}{r-3m}\right)^2 \sin U\cos V
 (d\theta^2+\sin^2\theta d\varphi^2) \ .
                                                \label{E3.6}
\end{equation}
It is straightforward to show that

1. $\Omega({\cal J}^+)=0$,

2. $\nabla\Omega({\cal J}^+)\not=0$ (namely,
   $\nabla_U \Omega=-\frac{1}{2\sqrt{27}m}=
    \nabla_V \Omega$ and
   $\nabla_\vartheta\Omega=0=
    \nabla_\varphi \Omega$ on ${\cal J}^+$),

3. $\hat g ^{\alpha\beta} \nabla_{\alpha}\Omega
    \nabla_{\beta}  \Omega\,\,({\cal J}^+)=-\Lambda/3<0$,

4. $d\hat s^2\,({\cal J}^+)=d U^2
  +\sin^2 U (d\theta^2+\sin^2\theta d\varphi^2)$.

We have thus demonstrated that $\Omega$ given by  (\ref{E3.5}) is
the proper conformal factor for the extreme Schwarzschild--de Sitter
space-time, and that its scri is smooth with geometry and topology
$S^3$, i.e. de Sitter-like.

\section {EMBEDDING DIAGRAM}

A natural time-slice in the extreme Schwarzschild--de Sitter
space-time (\ref{E2.1}) is $r=r_0=$ const.$\not=3m$, see Fig.~\ref{Fig.1}.
Introducing a coordinate $z$ by $z=\sqrt{-\Phi(r_0)}\, t$,
the metric of the slice is given by
\begin{equation}
ds^2\Big|_{r=r_0}=d z^2+r_0^2(d\theta^2+\sin^2\theta d\varphi^2) \ ,
     \label{E4.01}
\end{equation}
where $z\in (-\infty,+\infty)$, $\theta\in (0,\pi)$,
$\varphi\in (0,2\pi)$. Therefore, the embedding geometry is
a hyper-cylinder $R\times S^2$ with the two-spheres
having a constant radius $r_0$. Assuming, without loss of
generality, an equatorial section $\theta=\frac{\pi}{2}$, the embedding
diagram for $r=r_0$ (in any part of the complete space-time
between the points ${\cal P}$ for $r_0<3m$ or points ${\cal Q}$
for $r_0>3m$) is simply an infinite
cylinder with axis $z\sim t$ and radius $r_0$. This is clearly
similar to the embedding diagram of interior of the Schwarzschild
black hole ($r_0<2m$ for $\Lambda=0$).

Note also that the constant-mean-curvature foliation of the
Schwarzschild--de Sitter space-time has been found in
both extreme and non-extreme cases \cite{NMNO}. This different
time-slicing is particularly suitable for numerical studies
of gravitational collapse.

\section {SYNCHRONOUS COORDINATES}

In this section we shall present the synchronous (Lema\^\i tre-type)
coordinates for the extreme Schwarzschild--de Sitter space-time.
These coordinates connected with free particles moving radially
outward across the white-hole horizon towards infinity ($r=\infty$)
are useful for the discussion of the cosmic no-hair conjecture.

We introduce the coordinates $\tau$, $R$ by the relations
\begin{eqnarray}
d\tau&=&\quad dt-\frac{\sqrt{1-\Phi}}{\Phi} dr\ ,  \nonumber\\
dR   &=&-dt+\frac{1}{\Phi\sqrt{1-\Phi}}dr\ ,       \label{E4.1}
\end{eqnarray}
where $\Phi(r)$ is given by (\ref{E2.2}).
In these coordinates the metric (\ref{E2.1}) reads
\begin{equation}
ds^2=-d\tau^2+\left({\frac{2m}{r}+\frac{\Lambda}{3}\,r^2}\right)dR^2+
      r^2(d\theta^2+\sin^2\theta d\varphi^2) \ ,   \label{E4.2}
\end{equation}
with $m=1/3\sqrt{\Lambda}$ and
$r$ being given in term of $\tau$ and $R$ by
\begin{equation}
r(\tau, R)=\frac{2^{-1/3}}{\sqrt{\Lambda}\,Z}(1-Z^3)^{2/3} \ ,
\qquad
Z=e^{-\sqrt{\Lambda/3}\,(R+\tau)}                \ . \label{E4.5}
\end{equation}
The metric (\ref{E4.2}) is clearly regular at the horizon $r=3m$ where
$2m/r+(\Lambda/3)\,r^2=1$. Privileged timelike geodesic observers
with fixed $R=R_0$, $\theta=\theta_0$, $\varphi=\varphi_0$ start from
the past central singularity $r=0$ at $\tau=-R_0$ ($Z=1$)
and than move across the horizon
($Z^3=2-\sqrt 3$) to infinity $r=\infty$ as
$\tau\rightarrow\infty$ ($Z=0$).

One can easily bring the metric (\ref{E4.2}) into the form in
which the de Sitter metric in ``standard'' coordinates arises
explicitly. Introducing
$\chi=(2^{1/3}/\sqrt\Lambda)\exp({\sqrt{\Lambda/3}\,R})$,
the metric reads as follows:
\begin{equation}
ds^2=-d\tau^2+e^{2\sqrt{\Lambda/3}\,\tau}(1-Z^3)^{4/3}
    \left[{\left(\frac{1+Z^3}{1-Z^3}\right)^2 d\chi^2+
      \chi^2(d\theta^2+\sin^2\theta d\varphi^2)}\right] \ ,
\label{E4.7}
\end{equation}
where $Z$ is now given by
\begin{equation}
Z=\frac{2^{1/3}}{\sqrt\Lambda} \frac{e^{-\sqrt{\Lambda/3}\,\tau}}
     {\chi}\ . \label{E4.8}
\end{equation}
This is an exact form of the extreme Schwarzschild--de Sitter metric in
the outgoing comoving coordinates  describing a white hole
in the de Sitter universe. Keeping the leading order
terms in the expansion of (\ref{E4.7}) for $\tau\rightarrow\infty$, we obtain
\begin{eqnarray}
ds^2&=&-d\tau^2+e^{2\sqrt{\Lambda/3}\,\tau} \left[{ d\chi^2+
      \chi^2(d\theta^2+\sin^2\theta d\varphi^2)}\right] \label{E4.9}\\
    &&+\textstyle{\frac{2}{3}}(\sqrt{\Lambda}\,\chi)^{-3}
      e^{-\sqrt{\Lambda/3}\,\tau} \left[{ 2d\chi^2-
      \chi^2(d\theta^2+\sin^2\theta d\varphi^2)}\right]
      +{\cal O}\left(e^{-4\sqrt{\Lambda/3}\,\tau}\right) \ . \nonumber
\end{eqnarray}
The ``traces'' of the central white hole
completely disappear as $\tau\rightarrow\infty$ in full agreement with
the cosmic no-hair conjecture
--- asymptotically (near ${\cal J}^+$) we get the de Sitter
metric written in standard synchronous Friedmann-Robertson-Walker
form with the exponentially growing (``inflationary'') expansion factor.

\section {GEODESICS}

The metric of extreme Schwarzschild--de Sitter space-time
(\ref{E2.1}) is spherically symmetric. So we may, without
loss of generality, only consider geodesics which lie in a plane;
we choose $\theta=\frac{\pi}{2}$ here. Considering also the existence of
the Killing vectors $\partial_t$ and $\partial_\varphi$,
the geodesic equations can be written simply as
\begin{eqnarray}
\dot t       &=&\frac{E}{\Phi}\ ,  \label{E5.1a}\\
\dot \varphi &=&\frac{h}{r^2} \ ,  \label{E5.1b}\\
-\Phi\,{\dot t}^2+\Phi^{-1}\,{\dot r}^2
    +r^2\,{\dot\varphi}^2 &=&\epsilon\  , \label{E5.1c}
\end{eqnarray}
where $E$ and $h$ are constants, $\Phi$ is given by $(\ref{E2.2})$
and $\epsilon=-1, 0, +1$ for timelike, null and spacelike geodesics,
respectively. Here $\dot{\>}=d/d\tau$ where $\tau$ is a
(normalized) affine parameter along the geodesic. For
geodesics $r=$ const. we also have to consider the equation
\begin{equation}
\ddot r+\frac{1}{2}\Phi\Phi'\,{\dot t}^2-\frac{\Phi'}{2\Phi} \,{\dot r}^2
    -r\Phi\,{\dot\varphi}^2 =0\ , \label{E5.1d}
\end{equation}
($\>'=d/dr$) which for $r\not=$ const. follows from
$(\ref{E5.1a})$-$(\ref{E5.1c})$.

\subsection{Radial geodesics}

For radial motion $\varphi=\varphi_o=$ const. so that $h=0$.
Therefore, $(\ref{E5.1a})-(\ref{E5.1c})$ reduce to
\begin{eqnarray}
{\dot r}^2  &=&E^2+\epsilon\Phi\  , \label{E5.2a}\\
 \dot t     &=&\frac{E}{\Phi}\ .    \label{E5.2b}
\end{eqnarray}

\bigskip
\noindent
{\bf Null geodesics} $(\epsilon=0)$ can easily be integrated:
\begin{equation}
r(\tau)  =\pm |E|\,(\tau-\tau_0)  ,\quad
t(\tau)  =t_0\pm{\rm sgn} (E)\,r^*(r(\tau))\ , \label{E5.3}
\end{equation}
where $\tau_0$ and $t_0$ are constants, $r^*$ is given by
(\ref{E3.2}), and $E\not=0$ (for $E=0$
we get orbits which will be discussed bellow). Considering
$(\ref{E3.1})$, these null geodesics are given by
$\hat u=$ const. or $\hat v=$ const., i.e., they are indeed represented by
straight lines with inclination $45^0$ in the conformal diagram
in Fig.~\ref{Fig.1}.

\bigskip
\noindent
For {\bf timelike geodesics} $(\epsilon=-1)$ the right hand side
of $(\ref{E5.2a})$ can not be negative and the equation can be
integrated numerically. In Fig.~\ref{Fig.2} we show a typical free fall in
the black-hole space-time from $r_0>3m$ as a function the proper time $\tau$.
Observers with $E\not=0$ will reach the curvature singularity
at $r=0$ in a {\it finite} proper time whereas
special observers with $E=0$ will never cross the horizon $r=3m$.
Instead, they will approach the ``asymptotic points'' ${\cal P}$
indicated in Fig.~\ref{Fig.1}.
Indeed, for $E=0$ the equation of motion (\ref{E5.2a}) can
be integrated analytically (assuming $r(\tau_0)=r_0$):
\begin{eqnarray}
\tau-\tau_0&=&\int_r^{r_0} |\Phi(r)|^{-\frac{1}{2}}\,dr
     =3\sqrt{3}m \int_r^{r_0} \sqrt{\frac{r}{r+6m}}\frac{dr}{r-3m}
 \label{E5.4}\\
&=&3m\left[\sqrt{3}\ln(\sqrt{r^2+6mr}+r+3m)
+\ln\frac{|r-3m|}{\sqrt{3}\sqrt{r^2+6mr}+2r+3m}\right]_r^{r_0}
\ .\nonumber
\end{eqnarray}
Clearly, $r=\infty$ at $\tau=-\infty$, and for $r\to 3m$ we get
$\tau\approx-3m\ln|r-3m|\rightarrow\infty$.
Trajectories of timelike radial geodesic observers in the
conformal diagram are shown in Fig.~\ref{Fig.3}. The
trajectories in Fig.~\ref{Fig.3} (a) provide an
interpretation of the hypersurface $r=3m$ as
{\it an event horizon}  \cite{HE} for the  observers
with $E=0$. Also, the
expression $(\ref{E5.4})$ gives the proper
``distance'' of the horizon; it is infinite, as in the
case of the extreme Reissner-Nordstr\o m black hole and the extreme
Kerr black hole in asymptotically flat space-times (see, e.g.
\cite{Bi72}).

Note that the trajectories for $|E|=1$ can be found explicitly
--- they are given by $(\ref{E4.5})$. In fact,
these are the geodesics $R=R_0$, $\theta=\theta_0$,
$\varphi=\varphi_0$ in the synchronous coordinates $(\ref{E4.2})$
(or $\chi=\chi_0$ in the coordinates $(\ref{E4.7})$). Therefore,
the trajectories for $E=1$ shown in Fig.~\ref{Fig.3} (b) coincide
with the coordinate lines $R=R_0=$ const. in the conformal diagram.

\bigskip
\noindent
{\bf Spacelike geodesics} $(\epsilon=+1)$ are bounded for any
$E\not=0$, i.e. $r(\tau)$ is ``oscillating'' around horizons between
$r_{MIN}<3m<r_{MAX}$ for which the right hand side
of $(\ref{E5.2a})$ is positive. The result of a
numerical integration for different values of $E$ is presented in
Fig.~\ref{Fig.4} and a typical spacelike trajectory  in the
conformal diagram is shown in Fig.~\ref{Fig.5}. There are no analogues
of such geodesics in the Schwarzschild case (for
$\Lambda=0$). For $0<9\Lambda m^2\le1$ the geodesically complete
Schwarzschild--de Sitter space-time consists on an infinite number
of ``universes'' and black/white holes which are glued together
in the ``space-direction'' and thus ``tachyons'' can, in
principle, enter all of them.

\subsection{Nonradial geodesics}

For general geodesics $\varphi\not=$ const. ($h\not=0$) it is convenient to
rewrite Eq. (\ref{E5.1c}) by using (\ref{E5.1a}), (\ref{E5.1b})
and introducing the effective potential $V(r)$:
\begin{equation}
{\dot r}^2  =E^2-V(r)  , \label{E5.5a}
\end{equation}
where
\begin{equation}
 V(r)       =\Phi(r)\,\left(\frac{h^2}{r^2}-\epsilon\right)\ .
\label{E5.5b}
\end{equation}
Note however, that $(\ref{E5.5a})$-$(\ref{E5.5b})$ are {\it not}
equivalent to $(\ref{E5.1a})$-$(\ref{E5.1c})$ if $\Phi=0$, i.e.,
for $r=3m$; the corresponding motion related to the horizon must
be treated separately.

\bigskip
\noindent
For {\bf null geodesics} $(\epsilon=0)$ the effective potential
is shown in Fig.~\ref{Fig.6} (a). For all values of $h$ the potential grows
monotonically from $V=-\infty$ for $r=0$
to its maximal value $V=0$ at $r=3m$ and then decreases to
$V\rightarrow-h^2/27m^2$ asymptotically as $r\rightarrow+\infty$.
Therefore, qualitatively, as null particles with $E\not=0$
move from  $r=\infty$, they decelerate from the initial velocity
$|\dot r|=\sqrt{E^2+h^2/27m^2}$ to $|\dot r|=|E|>0$ at $r=3m$;
subsequently they accelerate so that $|\dot r|\rightarrow\infty$ as
$r\rightarrow0$. Starting from a finite $r$, the geodesics have
finite length before terminating at the physical singularity
$r=0$. This  behaviour can be  understood
considering the fact that in the region $r>3m$ the particle is
``pulled'' towards ${\cal J}$ by the $\Lambda>0$ term
whereas, for $r<3m$, the ``attractive'' influence of the black/white hole
($m>0$) becomes dominant. For $E=0$ the
geodesics equations can be integrated explicitly to give
\begin{equation}
\tau-\tau_0=\pm\frac{3\sqrt 3 m}{|h|}\left(
\sqrt{r^2+6mr}-3m\ln \left|6m\frac{\sqrt{3}\sqrt{r^2+6mr}+2r+3m}{r-3m}
\right|\right)  \ , \label{E5.5c}
\end{equation}
so that the horizon is not reached in a finite value of the
affine parameter ($\tau\rightarrow-\infty$ as $r\rightarrow 3m$)
in this case.

\bigskip
\noindent
For {\bf timelike geodesics} $(\epsilon=-1)$ the effective potential
resembles qualitatively the potential for $\epsilon=0$, except
that $V\rightarrow-\infty$ as $r\rightarrow+\infty$; it is shown in
Fig.~\ref{Fig.6} (b). Again, the particle's velocity reaches its
minimal value  $|\dot r|=|E|$ at $r=3m$. The conformal diagram
is not suitable for the visualization of nonradial trajectories since it
represents only the section $\theta=\theta_0$, $\varphi=\varphi_0$.
Therefore, in Fig.~\ref{Fig.7} we draw typical timelike nonradial geodesics
of freely falling particles with different nonvanishing $E$ and $h$
in the ``polar'' diagram $r(\varphi)$. Again,  the
singularity is reached in a finite proper time of any particle
with $E\not=0$. Geodesics with $E=0$ will be discussed in Section
8. Note that the nature of timelike
geodesics is much simpler in the extreme Schwarzschild--de
Sitter space-time if compared with non-extreme cases
$0<9m^2\Lambda<1$ (cf. \cite{Jak}) --- for example, there are no
bounded orbits.

\bigskip
\noindent
Nonradial {\bf spacelike geodesics} $(\epsilon=+1)$ in extreme
Schwarzschild--de Sitter space-time have a more complex structure
since the corresponding effective potential depends on $h$
in a nontrivial way. It is shown in Fig.~\ref{Fig.8}:
for all $h\not=0$, $V\rightarrow-\infty$ as $r\rightarrow 0$,
$V(3m)=0$ and $V\rightarrow+\infty$ as $r\rightarrow+\infty$.

For $0<|h|<3m$ the potential has a local minimum at $r_{min}=3m$
and a local maximum at $r_{max}$,  $0<r_{max}<3m$; it is given as
the unique solution of the equation
\begin{equation}
R^2(R^2+3R+9)=27H^2  , \label{E5.6}
\end{equation}
where $R=r/m$ and $H=h/m$. The dependence of $r_{max}$ and
$V(r_{max})$ on $h$ is shown in Fig.~\ref{Fig.9}
($r_{max}\approx \sqrt3 |h|$ for small $h$).
Therefore, for all $E^2<E_{max}^2=V(r_{max})$ there exist
bounded orbits (``oscillations'' around horizons) in the range
$r_{MIN}<3m<r_{MAX}$; typical bounded orbits in ``polar''
diagram are drawn in Fig.~\ref{Fig.10}.
 Of course, as one can see from the effective potential, there
are also geodesics in the range $0<r<r_{max}<3m$. For $E^2>E_{max}^2$
there exist a maximum value of $r$ for all the geodesics which
necessarily  reach the singularity at $r=0$. Special geodesics
$|E|=|E_{max}|$ will be discussed in Section~8.

For $|h|=3m$ the potential has a point of inflexion at $r_{min}=3m$.
Qualitatively, such spacelike ``observers'' with $E\not=0$ fall from
a maximum value of $r$ to the singularity with a growing velocity.
Similar motions are found for $|h|>3m$ with a difference that the ``observers''
decelerate in the range $3m<r<r_{min}$ since the potential now has a local
maximum $V=0$ at $r=3m$ and a local minimum $V<0$ at $r_{min}$ which
is the solution of $(\ref{E5.6})$. Geodesics with $E=0$
approaching the ``asymptotic points'' ${\cal P, Q}$ we analyze in
Section 8.

\subsection{Circular orbits}

In this section we shall finally establish special nonradial
geodesics in the extreme Schwarz\-schild-de Sitter space-time ---
circular orbits $r=r_0=$ const. The effective potential
analysis presented above indicates that such orbits may exist
only in the extreme of $V(r)\ge 0$, i.e. for $\epsilon=+1$,
$0<|h|<3m$. However, since the case $V=0$ must be treated
separately, it is better to return back to $(\ref{E5.1a})$-$(\ref{E5.1c})$
and $(\ref{E5.1d})$.

For $r=r_0\not=3m$ these equations imply
\begin{eqnarray}
\Phi_0\left(\frac{h^2}{r_0^2}-\epsilon\right) &=&E^2\  ,\quad
\epsilon r_0^2\left(1-\frac{2\Phi_0}{r_0\Phi'_0}\right)^{-1} \ =\
 h^2\  ,\nonumber\\
\dot t       & =& \ \frac{E}{\Phi_0}\ ,  \quad
\dot \varphi \ =\ \frac{h}{r_0^2} \ . \label{E5.7}
\end{eqnarray}
It can now be shown that there are no
circular null or timelike geodesics of this type; there is a unique (unstable)
circular spacelike geodesic $0<r_0<3m$ for any $0<|h|<3m$ corresponding
to the local maximum of the effective potential $r_{max}$ given by the
solution of $(\ref{E5.6})$.

For $r=r_0=3m$ only Eqs. $(\ref{E5.1b})$, $(\ref{E5.1c})$
are nontrivial yielding $r_0^2{\dot\varphi}^2=\epsilon$ and
$\dot\varphi=h/r_0$, respectively, so that $h^2=\epsilon
r_0^2=9m^2\epsilon$. Therefore, there is only one null circular geodesic
$r=3m$ (for $h=0$) and one spacelike circular orbit (for
$|h|=3m$). Of course, the same result we get if we
start from the Kruskal-type coordinates $(\ref{E3.3})$.

To summarize, there are {\it no timelike circular geodesics} in
the extreme Schwarzschild--de Sitter spacetime, and there is a {\it unique
null} circular geodesic $r=3m$ (with $E=0=h$) --- the horizon
(cf. \cite{Stu}). For any $0< |h|\le 3m$ there  exists a unique {\it unstable
spacelike} circular orbit $0<r_0\le3m$. These results contrast
with those for non-extreme cases  $0<9\Lambda m^2<1$ (cf.
\cite{Jak}), or for the $\Lambda=0$ case \cite{MTW}. For
example, in the Schwarzschild space-time the circular
photon orbit at $r=3m$ is situated outside the horizon $r=2m$.

Note that throughout this section we have used the terms
``bounded'' or ``circular'' orbit in connection with the coordinate
$r$ although in fact it is a {\it time coordinate}. However, it is
reasonable to keep such a description since $r$ still measures
the ``distance from the singularity'' situated at $r=0$. We could
alternatively associate these terms  with the space coordinate
$t$, but using Eq. $(\ref{E5.1a})$ and considering $\Phi\leq0$
we see that $t$ is always a {\it monotone} function of $\tau$.
Only for $E=0$ the geodesics are ``t-circular''.

\section {GEODESIC DEVIATION}

Let us consider an arbitrary radial timelike observer in
the extreme Schwarzschild--de Sitter space-time (\ref{E2.1}),
(\ref{E2.2}). We can set up an
orthonormal parallelly propagated frame
\begin{equation}
{\vec{\bf e}}_{(0)} =\dot t\partial_t+\dot r\partial_r
     \ ,\quad
{\vec{\bf e}}_{(1)}   =\frac{1}{r}\partial_\theta
     \ ,\quad
{\vec{\bf e}}_{(2)} =\frac{1}{r\sin\theta}\partial_\varphi
     \ ,\quad
{\vec{\bf e}}_{(3)}   =\frac{\dot r}{\Phi}\partial_t+\Phi\dot t\partial_r\
, \label{E6.1}
\end{equation}
where ${\vec{\bf e}}_{(0)}$ is the four-velocity of the observer;
all the coefficients must be evaluated at a given proper
time $\tau$ of the geodesic observer, $r=r(\tau)$ etc. Projecting
the curvature tensor of the space-time onto the frame (\ref{E6.1}) we get
a coordinate-independent form of the equation of geodesic deviation
\begin{eqnarray}
{\ddot Z}^{(1)}  &=&\frac{\Lambda}{3} Z^{(1)}-\frac{m}{r^3} Z^{(1)}
     \ ,\nonumber \\
{\ddot Z}^{(2)}  &=&\frac{\Lambda}{3} Z^{(2)}-\frac{m}{r^3} Z^{(2)}
     \  ,\label{E6.2}  \\
{\ddot Z}^{(3)}  &=&\frac{\Lambda}{3} Z^{(3)}+\frac{2m}{r^3} Z^{(3)}
     \ ,\nonumber
\end{eqnarray}
where $Z^{(i)}=Z^\mu e^{(i)}_\mu$, $i=1, 2,3$, are frame components of
the vector connecting two nearby free test particles. For $r\rightarrow\infty$
the cosmological constant $\Lambda$ dominates in (\ref{E6.2}) so that
asymptotically $Z^{(i)}\approx\exp(\sqrt{\frac{\Lambda}{3}}\,\tau)$
as the observers approach exponentially expanding de Sitter-like
infinity (cf. (\ref{E4.9})). On the other hand, falling to the singularity at
$r=0$, the $\Lambda$-terms become negligible; the observers
are stretched by tidal forces in the radial direction and are
squeezed in the perpendicular directions as in the
Schwarzschild black-hole space-time.

Exact solutions  can be obtained numerically by a simultaneous
integration of (\ref{E5.2a}) and (\ref{E6.2}).
In this section, however,  we concentrate on analytic investigation
of the behaviour of particles approaching the asymptotic regions
given by the ``points''
${\cal P}$ and ${\cal Q}$ along timelike radial geodesics with
$E=0$. These are given explicitly by (\ref{E5.4}) so that
$r\approx3m+A\exp(-\tau/3m)$ as $\tau\rightarrow\infty$,
where $A$ is a constant (positive  if approaching ${\cal P}$ and negative
if approaching ${\cal Q}$). The asymptotic form of
(\ref{E6.2}) is then
\begin{eqnarray}
{\ddot Z}^{(j)}  &=&a\,e^{-\tau/3m} Z^{(j)}\ ,\nonumber \\
{\ddot Z}^{(3)}  &=&\left(\Lambda-2a\,e^{-\tau/3m}\right) Z^{(3)}
     \ , \label{E6.3}
\end{eqnarray}
where $j=1, 2$ and $a=A\Lambda/3m$. Performing substitutions
\begin{eqnarray}
T         &=&6m\sqrt{|a|}\,  e^{-\tau/6m} \ ,\nonumber \\
\tilde T  &=&6m\sqrt{|2a|}\, e^{-\tau/6m} \ , \label{E6.4}
\end{eqnarray}
these equations go over to the Bessel equation so that general
solutions of (\ref{E6.3}) are
\begin{eqnarray}
Z^{(j)}(T)        &=&A_j I_0(T)+B_j K_0(T)\ ,\nonumber \\
Z^{(3)}(\tilde T) &=&C J_2(\tilde T)+D N_2(\tilde T) \ , \label{E6.5}
\end{eqnarray}
for particles approaching ${\cal P}$ and
\begin{eqnarray}
Z^{(j)}(T)        &=&A_j J_0(T)+B_j N_0(T)\ ,\nonumber \\
Z^{(3)}(\tilde T) &=&C  I_2(\tilde T)+D K_2(\tilde T) \ , \label{E6.6}
\end{eqnarray}
for those approaching ${\cal Q}$, where $A_j, B_j, C, D$ are constants.
Expansions of (\ref{E6.5}) and (\ref{E6.6}) for
$\tau\rightarrow\infty$ give
\begin{eqnarray}
Z^{(j)} &\approx& \alpha_j +\beta_j \tau+
     \frac{\beta_j A}{3m}\tau e^{-\tau/3m}+\cdots \ ,\label{E6.7}\\
Z^{(3)} &\approx& \gamma  e^{-\tau/3m}
     \left(1- \frac{2A}{9m}e^{-\tau/3m}+\cdots\right)+
               \delta  e^{\tau/3m}
     \left(1+ \frac{2A}{9m}e^{-\tau/3m}+\cdots\right) \ ,\nonumber
\end{eqnarray}
where $\alpha_j, \beta_j, \gamma, \delta$ are constants. Therefore,
relative motion in the perpendicular directions ${\vec{\bf
e}}_{(1)}$, ${\vec{\bf e}}_{(2)}$ is {\it uniform} as
$\tau\rightarrow\infty$; for a special choice of initial
conditions, $\beta_j=0$, we get $Z^{(j)}\rightarrow$ const.
Similarly, for $\delta=0$ the motion in the radial direction
${\vec{\bf e}}_{(3)}$ is given by $Z^{(3)}\rightarrow 0$. Thus,
relative motion close to ${\cal P}$ and ${\cal Q}$ is {\it nonsingular}.
This supports our physical interpretation of ${\cal P}$ and ${\cal Q}$ as
``asymptotic points'' representing regions which do {\it not} belong
to the singularity at $r=0$ or to the de Sitter-like infinity $r=\infty$,
although they seem to ``lie'' on the same lines in the conformal diagram
shown in Fig.~\ref{Fig.1}.

\section {NATURE OF THE ``ASYMPTOTIC POINTS'' ${\cal P, Q}$}

In this section we shall look at these regions represented by
``points'' ${\cal P, Q}$ more closely.
The ``points'' ${\cal P}$ in the conformal diagram are given by
$u=-\infty$, $v=+\infty$ whereas  ${\cal  Q}$ are given by
$u=+\infty$, $v=-\infty$. It was demonstrated in the previous section
that relative motions of observers approaching
${\cal P}$, ${\cal Q}$ differ significantly from those
corresponding to $r\rightarrow0$ or $r\rightarrow\infty$. In
fact, ${\cal P}$ and ${\cal Q}$  represent asymptotic regions
which are reached by a family of special observers such
that $r\rightarrow3m$ as $\tau\rightarrow\infty$; observers from
$r\geq3m$ reach ${\cal P}$, and observers from $r\leq3m$ reach
${\cal Q}$. Since $V(r\rightarrow3m)\rightarrow0$, we see
from   (\ref{E5.5a}) that these geodesic observers must have $E=0$
(otherwise they would reach $r=3m$ in a finite $\tau$).
Eq. (\ref{E5.1a}) then gives $t=t_0=$ const. (note again that $t$
is a space coordinate) and  the
trajectories of all such observers in the conformal diagram
coincide with those presented in Fig.~\ref{Fig.3}  (a).
The effective potentials shown in Figs.~\ref{Fig.6} (a),  (b) and
Fig.~\ref{Fig.8}  for
$\epsilon=0, -1, +1$, respectively, indicate that all geodesic
observers  with $E=0$ can reach $r=3m$ asymptotically
except for spacelike radial observers ($\epsilon=+1, h=0$).
Therefore, the family of geodesics approaching ${\cal P}$
consists of the following:

\begin{enumerate}
\item {\it null radial} ($\epsilon=0, h=0$); this is the circular orbit
on the horizon (see Section 6.3).

\item{\it timelike radial} ($\epsilon=-1, h=0$); these are given by
(\ref{E5.4}) so that $r^*\approx 1/(r-3m)\approx \exp(\tau/3m)$, i.e.,
$u=t_0-r^*\rightarrow-\infty$, $v=t_0+r^*\rightarrow+\infty$
as $\tau\rightarrow\infty$.

\item{\it null nonradial} ($\epsilon=0, h\not=0$); they are given by
Eq. (\ref{E5.5c}) so that $r^*\approx  \exp(C_1\tau)$, where
$C_1=|h|/9\sqrt3 m^2$, implying
$u=\rightarrow-\infty$, $v\rightarrow+\infty$.

\item{\it timelike nonradial} ($\epsilon=-1, h\not=0$);
introducing $\xi\equiv r-3m$ the Eqs. (\ref{E5.5a}),
(\ref{E5.5b}) can be written for $r\rightarrow3m$ as
$\dot\xi\approx -C_2\xi$, where $C_2=\sqrt{9m^2+h^2}/9m^2$. Therefore,
$r^*\approx 1/\xi\approx \exp(C_2\tau)\rightarrow\infty$ so that
$u=\rightarrow-\infty$, $v\rightarrow+\infty$.

\item{\it spacelike nonradial} ($\epsilon=+1, h\not=0$);
from Fig.~\ref{Fig.8} it follows that geodesics of this type with $E=0$
and $r\geq3m$ can exist only if $|h|\geq3m$. As in the case 4.,
for  $r\rightarrow3m$  we get
$r^*\approx 1/\xi\approx \exp(C_3\tau)\rightarrow\infty$, where
$C_3=\sqrt{h^2-9m^2}/9m^2$, i.e., $u=\rightarrow-\infty$,
$v\rightarrow+\infty$ as $\tau\rightarrow\infty$. Note that
trajectories of such observers in the conformal diagram (in contrast
to all the previous cases) do not extend to $r=\infty$. Instead,
they ``make loops'' around ${\cal P}$: tachyons moving ``outward''
would reach the maximum value of $r$ (which is $r=|h|$) and then
they approach the same ${\cal P}$ again asymptotically as
$\tau\rightarrow\infty$.
\end{enumerate}

\noindent
Similarly, the geodesics approaching ${\cal Q}$ are:

\begin{enumerate}
\item {\it null radial} ; this is the circular orbit
on the horizon.

\item{\it timelike radial} ; from (\ref{E5.4}) it follows that
$r^*\approx 1/(r-3m)\approx -\exp(\tau/3m)$, i.e.,
$u=\rightarrow+\infty$, $v\rightarrow-\infty$
as $\tau\rightarrow\infty$.

\item{\it null nonradial} ; Eq. (\ref{E5.5c}) gives
$r^*\approx  -\exp(C_1\tau)$ so that
$u=\rightarrow-\infty$, $v\rightarrow+\infty$.

\item{\it timelike nonradial} ; Eqs.
(\ref{E5.5a}) and (\ref{E5.5b}) for $\eta\equiv 3m-r$ give
$\dot\eta\approx -C_2\eta$ so that
$r^*\approx -1/\eta\approx -\exp(C_2\tau)\rightarrow\infty$. Again,
$u=\rightarrow-\infty$, $v\rightarrow+\infty$ as
$\tau\rightarrow\infty$.

\item{\it spacelike nonradial} ;
these geodesics exist for $|h|\geq3m$ only. If $|h|>3m$ we get
$r^*\approx -\exp(C_3\tau)\rightarrow-\infty$.
If $|h|=3m$ the geodesics with $E=0$ are approaching the point of
inflexion of $V(r)$ at $r=3m$ and in such a case Eqs. (\ref{E5.5a})
and (\ref{E5.5b}) give $\dot\eta\approx\sqrt{2/27m^3}\eta^{3/2}$
so that $r^*\approx -\tau^2\rightarrow-\infty$.
In both cases,
$u=\rightarrow-\infty$, $v\rightarrow+\infty$.

\end{enumerate}

\noindent
It can be shown that asymptotic motion such that $r \rightarrow
r_0=$ const. as $\tau\rightarrow\infty$ is possible only if
$E=V(r_0)$ where $r_0$ is either a {\it local maximum} or a {\it
point of inflexion} of the effective potential. Therefore,  the above
geodesics represent the only geodesic motion approaching
$r=3m$ asymptotically, i.e. regions ${\cal P}$ and ${\cal Q}$.

Note also that for $h$ such that  $0<|h|<3m$ there  exists a special
class of nonradial spacelike geodesics with $|E|=|E_{max}|$ (cf.
Sec. 6.2) which asymptotically approach $r_{max}$ as
$\tau\rightarrow\infty$. Here, $r_{max}$ depends on $h$, as
indicated in Fig.~\ref{Fig.9}, and represents the local maximum of $V(r)$,
$0<r_{max}<3m$. Eq. (\ref{E5.2b}) gives
$t\approx\pm(|E_{max}|/\Phi(r_{max}))\tau\rightarrow\pm\infty$,
i.e., these geodesics also
seem to approach  ${\cal P, Q}$ in the conformal
diagram. However, they converge to $r_{max}$ which is different
from $r=3m$ corresponding to ${\cal P, Q}$. In fact,
$u=t-r^*(r_{max})\rightarrow\pm\infty$ and
$v=t+r^*(r_{max})\rightarrow\pm\infty$.

\section {REDSHIFT}

Finally, we shall investigate the redshift of signals emitted by a
source falling into the singularity $r=0$ in the extreme Schwarzschild--de
Sitter space-time representing a black hole, see Fig.~\ref{Fig.1} (a).
We shall assume that the source follows a timelike radial
geodesic ($\epsilon=-1$, $h=0$) given by Eqs. (\ref{E5.2a}) and
(\ref{E5.2b}) with $E>0$. Geodesics of this type are shown in
Fig.~\ref{Fig.3} (b); they start in the region $r>3m$ and
reach the horizon $r=3m$, $t=-\infty$ in a finite proper time.
At an event $(t_e, r_e>3m)$ the source emits a signal with frequency
$\omega_e$ which propagates along the null radial geodesic $\hat u=$
const. given by Eq. (\ref{E5.3}). Our goal here is to investigate
the redshift $z=\omega_o/\omega_e-1$, where $\omega_o$ is the
frequency of the signal as measured by an observer remaining outside the
black-hole horizon. In asymptotically flat black-hole space-times
it is standard to choose static distant outer observer,
$r=$ const.$\to\infty$.
Unfortunately, for principal reasons we can not make
this natural choice here. The extreme Schwarzschild--de Sitter space-time
is not asymptotically flat and contains no static region
($r$ is a time coordinate, and $r=\infty$ represents
${\cal J}^-$, i.e. time and null past infinity).
Moreover, as we have observed in Section 6, most observers
starting at $r>3m$ cross the horizon and fall
into the singularity. The only ``reasonable'' outer
observers are those approaching the nonsingular asymptotic
region given by ${\cal P}$. Therefore, we shall assume that $\omega_o$
is detected by an observer moving along a timelike radial
geodesic with $E=0$ given $\dot r_o=-\sqrt{-\Phi(r_o)}$, $t_o=$
const., with $\tau$ being observer's proper time (see Eqs.
(\ref{E5.2a}), (\ref{E5.2b})); typical trajectories of this type
are shown in Fig.~\ref{Fig.3}~(a).

It is now straightforward to calculate the redshift using the
well known formula $z=(k_\mu u^\mu_e)_e/(k_\mu u^\mu_o)_o-1$,
where $u^\mu_e=(E/\Phi(r_e), -\sqrt{E^2-\Phi(r_e)}, 0, 0)$ is the
four-velocity of the source, $u^\mu_o=(0, -\sqrt{-\Phi(r_o)}, 0, 0)$ is the
four-velocity of the observer, and $k^\mu=(-1/\Phi(r), -1, 0, 0)$ is the
null vector tangent to the photon trajectory; we get
\begin{equation}
 z =\sqrt{-\Phi(r_o)}\,\,\frac{E+\sqrt{E^2-\Phi(r_e)}}{-\Phi(r_e)}-1\ .
\label{E10.1}
\end{equation}
For a source approaching the horizon,
$r_e\to3m$, we have also $r_o\to3m$ (since $3m<r_o<r_e$).
Near the horizon, $\Phi(r)$ given by
(\ref{E2.2}) can be written as $\Phi(r)\approx-(r-3m)^2/9m^2$, so
that we can express (\ref{E10.1}) in the  form
$z\approx 6mE\,(r_o-3m)/(r_e-3m)^2-1$. It only remains to find a
relation between $r_e$ and $r_o$. Since the emission and
observation events are connected by a photon trajectory $u=$
const., we have $t_e-r^*(r_e)=t_o-r^*(r_o)$, where $r^*$ is given
by (\ref{E3.2}). The relation between $t_e$ and $r_e$ follows
from Eqs. (\ref{E5.2a}) and (\ref{E5.2b}), $t_e+\hbox{const.}
=-E\int[\Phi(r_e)\sqrt{E^2-\Phi(r_e)}]^{-1}\,dr_e
\approx-\int\Phi^{-1}(r_e)\,dr=-r^*(r_e)$. Thus,
$2r^*(r_e)=r^*(r_o)+$const., implying $(r_e-3m)\approx2\,(r_o-3m)$.
Considering finally $r_o-3m\approx\exp(-\tau/3m)$ (see
(\ref{E5.4})), we arrive at the formula
\begin{equation}
 z \approx \exp{\left(\frac{\tau}{3m}\right)}\ .
\label{E10.2}
\end{equation}
The redshift grows exponentially with the characteristic
$e$-folding time $\tau_e=3m=1/\sqrt{\Lambda}$. This result may
seem somewhat surprising since, for extreme Reissner-Nordstr\o m
and extreme Kerr black holes, the redshifts are given by
power laws \cite{Bi72}, \cite{BS}. However, we should emphasize
again that these redshifts were calculated with respect to
distant observers ``$r=\infty$'' in asymptotically flat
solutions, contrary to our case where $r\to 3m$ for observers
approaching the point ${\cal P}$ in non-asymptotically
flat extreme Schwarzschild--de Sitter black hole space-time.

\section {CONCLUSION}

We have analyzed the extreme Schwarzschild--de Sitter
space-time describing a spherically symmetric black (or white)
holes in the de Sitter universe characterized by the condition
$9\Lambda m^2=1$. Coordinates suitable for rigorous discussion of
the global structure and the cosmic no-hair conjecture have been
introduced. All possible geodesic motions have also been
investigated and, with the help of the equation of geodesic
deviation, the nature of specific nonsingular ``asymptotic
points'' ${\cal P, Q}$ in the conformal diagram has been studied.
It has been demonstrated that they represent whole asymptotic
regions for large classes of geodesics (${\cal P}$ separate
singularities of different black/white holes and ${\cal Q}$
separate different de Sitter-like past/future infinites).
Observers approaching these regions radially detect exponentially
growing redshift of signals emitted by particles falling to the
singularity.

\section {ACKNOWLEDGMENTS}

Author thanks Ji\v r\'\i\ Bi\v c\' ak for many very stimulating
discussions and Jerry Griffiths for reading the manuscript. The
support of grants Nos. GACR-202/99/0261 and
GAUK-230/1996 from the Czech Republic and Charles University is
also acknowledged.

\newpage

\begin{figure}
\caption{
Conformal diagram of the extreme Schwarzschild--de
Sitter space-time with $9\Lambda m^2=1$. (a) The singularity $r=0$
in future, corresponding to black holes. The maximal analytic
extension of the geometry is obtained by glueing an infinite
number of regions shown in the figure, or joining a finite number of
regions via identification of events along two horizons $r=3m$.
(b) The time-reversed diagram ($\hat u\rightarrow-\hat u, \hat v
\rightarrow-\hat v$), corresponding to white holes.}
\label{Fig.1}
\end{figure}

\begin{figure}
\caption{
A geodesic timelike observer falling radially
will reach the horizon $r=3m$ and the black-hole singularity
$r=0$ in a finite proper time $\tau$. Only if the constant of motion $E$
vanishes, $\tau\rightarrow\infty$ as $r\rightarrow 3m$.
Here we assume $r=r_0=10m$ at $\tau=0$.}
\label{Fig.2}
\end{figure}

\begin{figure}
\caption{
Trajectories of typical timelike radial geodesics in the
conformal diagram for observers:
(a) with $E=0$ which are approaching asymptotic points ${\cal P, Q}$,
(b) with $E\not=0$ (here we assume $E=1$).}
\label{Fig.3}
\end{figure}

\begin{figure}
\caption{
Spacelike radial geodesics oscillate around $r=3m$ corresponding
to different horizons. The amplitude of oscillations grows with
a growing value of $E$.}
\label{Fig.4}
\end{figure}

\begin{figure}
\caption{
Typical trajectory of a spacelike radial geodesic in the
conformal diagram of the extreme Schwarzschild-de Sitter space-time
(here we assume $E=1$).}
\label{Fig.5}
\end{figure}

\begin{figure}
\caption{
Effective potentials $V(r)$ for nonradial motions in
the extreme Schwarzschild--de Sitter space-time for
(a) null geodesics, (b) timelike geodesics.}
\label{Fig.6}
\end{figure}

\begin{figure}
\caption{
Typical timelike nonradial geodesics drawn
in the ``polar'' diagram, $r(\varphi)$, for different values of
$E$, (a) for $h=0.2 m$, (b) for $h=m$. Starting from $r=10 m$ they cross
the horizon $r=3m$ and reach the singulaity $r=0$ in a finite value
of the proper time.}
\label{Fig.7}
\end{figure}

\begin{figure}
\caption{
Effective potentials $V(r)$ for nonradial spacelike
geodesics depend significantly on $h$. See the text for more
details.}
\label{Fig.8}
\end{figure}

\begin{figure}
\caption{
Plots of the local maximum $r_{max}$ of the
effective potential and $V(r_{max})$ for nonradial spacelike geodesics with
$0<|h|<3m$ as a function of $h$.}
\label{Fig.9}
\end{figure}

\begin{figure}
\caption{
Typical spacelike nonradial geodesics drawn
in the ``polar'' diagram, $r(\varphi)$, for different values of
$E$, (a) for $h=0.2 m$, (b) for $h=0.5 m$.}
\label{Fig.10}
\end{figure}

\end{document}